\documentclass[aps,pra,showpacs,twocolumn, longbibliography]{revtex4-1} 

\usepackage{indentfirst}
\usepackage{graphicx}
\usepackage{dcolumn}
\usepackage{bm}
\usepackage{color}
\usepackage{amsmath}
\usepackage{amssymb}
\usepackage{adjustbox}
\usepackage{array}

\newcommand{\tabincell}[2]{\begin{tabular}{@{}#1@{}}#2\end{tabular}}

\begin{document}

\title{Broadband-laser-diode pumped PPKTP-Sagnac polarization-entangled photon source}

\author{Neng Cai$^{1, 4}$}
%
\author{Wu-Hao Cai$^{1, 4}$}
%
\author{Shun Wang$^{1}$}
%
\author{Fang Li$^{1}$}
\email{lifang@wit.edu.cn}
\author{Ryosuke Shimizu$^{2}$}
\email{r-simizu@uec.ac.jp}
\author{Rui-Bo Jin$^{1, 3}$}
\email{jrbqyj@gmail.com}

\affiliation{$^{1}$ Hubei Key Laboratory of Optical Information and  Pattern Recognition, Wuhan Institute of Technology, Wuhan 430205, China}
\affiliation{$^{2}$ The University of Electro-Communications, 1-5-1 Chofugaoka, Chofu, Tokyo, Japan}
\affiliation{$^{3}$ Guangdong Provincial Key Laboratory of Quantum Science and Engineering, Southern University of Science and Technology, Shenzhen 518055, China}
\affiliation{$^{4}$ These authors contributed equally to this work}


%


\date{\today }
\begin{abstract}
We experimentally demonstrate a polarization-entangled photon source at 810 nm using a type-II phase-matched PPKTP crystal pumped by a low-cost, broadband laser diode with a central wavelength of 405 nm and a typical bandwidth of 0.53 nm.
The PPKTP crystal is placed in a Sagnac-loop to realize the compact size and high stability.
The downconverted biphotons, the signal and the idler, have typical bandwidths of 5.57 nm and 7.32 nm.
We prepare two Bell states $|\Psi^+\rangle$ and $|\Psi^-\rangle$ with the fidelities of 0.948$\pm$0.004 and 0.963$\pm$0.002.
In polarization correlation measurement, the visibilities are all higher than 96.2\%, and in the Bell inequality test, the S value can achieve 2.78$\pm$0.01.
This high-quality and low-cost entangled photon source may have many practical applications in quantum information processing.
\end{abstract}


\maketitle

\section{Introduction}
Quantum entanglement, which represents a non-classical correlation among several quantum subsystems, is a critical feature of quantum information science.
High-quality entangled photon pairs play an important role in many quantum information technologies, such as quantum communication \cite{Vaziri2002,Boileau2004}, quantum computation \cite{Browne2005,Walther2005}, and quantum measurement  \cite{Maccone2020PRL}.
A photon has many degrees of freedom, e.g., time, frequency, polarization, position, momentum, and each degree of freedom can be entangled \cite{Kuzucu2005, Oohata2007, Howell2004, Romero2012}.
Polarization-entangled photons, which are relatively easy to generate and characterize, have been widely investigated.
As early as 1988, Shih \emph{et al.} \cite{Shih1988} and Ou \emph{et al.} \cite{Ou1988} demonstrated the polarization-entangled photons using Type-I phase-matched spontaneous parametric down-conversion (SPDC) scheme by pumping potassium dihydrogen phosphate (KDP) crystals with  monochromatic continuous-wave (CW) lasers.
In 1995,  Kwiat \emph{et al.} showed a new entangled photon source by pumping a $\beta$-barium borate (BBO) crystal under type-II phase-matching condition using argon-ion laser \cite{Kwiat1995}.
From then on, many schemes of entangled photon sources  have been demonstrated with different pump lasers (monochromatic CW laser \cite{Grice1997,Li2015} or ultra-fast pulsed laser \cite{Keller1997,Kim2019}), different nonlinear crystals (Periodically polarized potassium titania phosphate (PPKTP) \cite{Weston2016}, BBO \cite{Kwiat1999}, or periodically poled lithium niobate (PPLN) \cite{Koenig2005}), different phase-matching types (Type-0 \cite{ChenYY2018PRL, Terashima2018}, Type-I \cite{Rangarajan2009, Villar2018},  or Type II \cite{Martin2010,Horn2019,Lee2016}),  or different wavelengths (e.g., 810 nm  or 1550 nm). See a recent review article on entangled photon sources in \cite{Anwar2021}.

It can be noticed from the above works that the pump lasers used in the previous entangled sources are almost monochromatic CW lasers or ultrafast pulsed lasers.
However, the low-cost broadband multi-mode (longitudinal mode) laser is not widely used.
Recently, quantum optical technologies are spreading out from laboratory to industrialization.
In practical use, the low-cost and high stability entangled photon sources are indispensable.
With the rapid development of blue laser technologies, the inexpensive high-power blue laser diode (LD) can easily provide high power of over 100 mW at 405 nm.
Therefore, it is necessary to investigate the multi-mode laser, especially to apply it to the field of quantum optics, e.g., for the preparation of an entangled photon source.
Recently,  Jeong \emph{et al.} adopted the method of ``universal Bell-state synthesizer'' \cite{Kim2003} to prepare an entangled photon source by pumping a type-II phase-matched PPKTP with a broadband multi-mode LD \cite{Jeong2016}.
Lohrmann \emph{et al.} prepared an entangled photon-pair source using the configuration of ``linear beam displacement interferometer'' using broadband LD and type-0 phase-matched PPKTP \cite{Lohrmann2020APL}.

Compared with the above two schemes, there are other optical path configurations, such as the Sagnac-loop structure, which has the merits of compact design and high stability.
The first polarization-entangled photon pair using a Sagnac interferometer was demonstrated by Shi \emph{et al.} \cite{Shi2004} with a BBO crystal in 2004.
Later, Kim \emph{et al.} optimized the scheme by using a type-II phase-matched PPKTP crystal for higher brightness and stability \cite{Kim2006}.
Subsequently, the Sagnac-PPKTP scheme has been widely used in the research of quantum entangled sources at around 800 nm wavelength.
For example, a bright entangled photon source was realized by using a mode-locked pulse pumping \cite{Kuzucu2008}; a wavelength-tunable and narrow-band entangled photon source was demonstrated by using a CW pumping \cite{Fedrizzi2007};  a non-collinear PPKTP-Sagnac scheme was demonstrated using a type-0 phase-matched PPKTP \cite{Jabir2017}. Especially, a high-quality satellite-based entangled source adopted the configuration of the PPKTP-Sagnac scheme and pumped by single-longitudinal-mode LD \cite{Yin2017, Zhou2015JIMW}.
The PPKTP-Sagnac entangled photon source was also demonstrated at telecom wavelengths and pumped by pico-second laser \cite{Jin2014OE}, femto-second  laser \cite{Weston2016}, or CW lasers \cite{Li2015}. The PPKTP-Sagnac scheme at the telecom wavelengths has  extra merit of high spectral purity \cite{Jin2013OE}.
In this work, we further develop the PPKTP-Sagnac scheme to be pumped by a broadband multi-mode LD. This new polarization-entangled photon source has the merits of low-cost, high brightness, and high stability.

\section{Experiment}
\begin{figure}[tbh]
\centering\includegraphics[width= 0.50\textwidth]{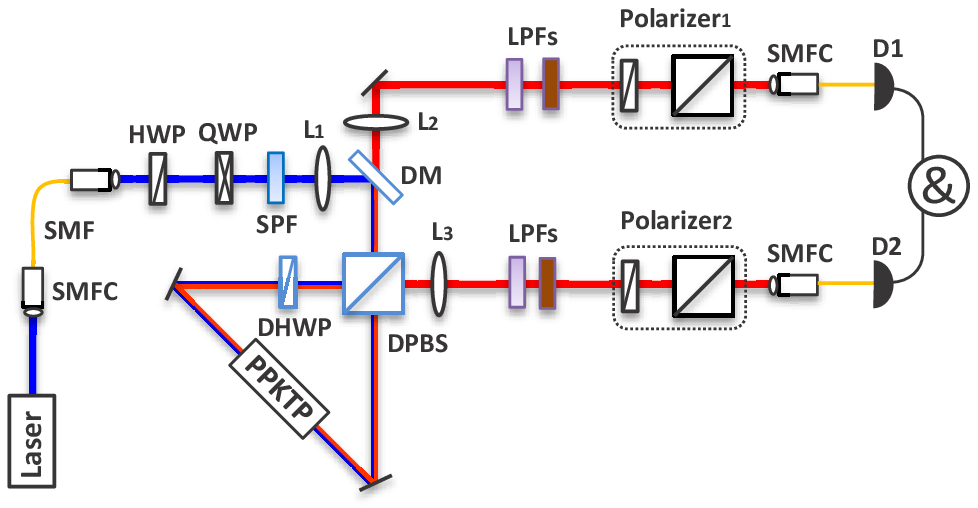}
\caption{Experimental setup for the LD pumped type-II-PPKTP-Sagnac entangled photon source. SMFC=single-mode fiber coupler, SMF=single-mode fiber,  HWP=half-wave plate, QWP=quarter-wave plate, SPF=short-pass filter, L=lens, DM=dichroic mirror, PBS=polarization beam splitter, DPBS=dual-wavelength PBS, DHWP=dual-wavelength HWP, LPF=long-wave pass filters, D=detector, $\&$=coincidence counter.
 } \label{fig:1}
\end{figure}
\begin{figure}[htb]
\centering\includegraphics[width= 0.48 \textwidth]{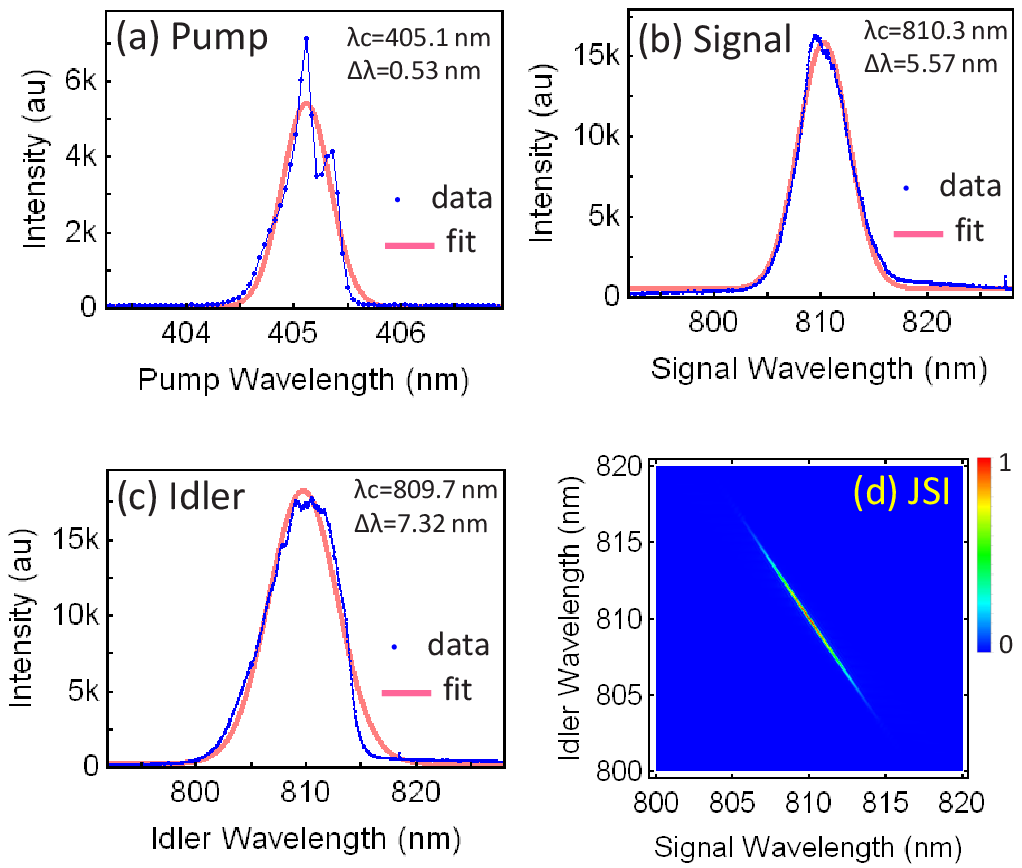}
\caption{(a) is a typical spectrum of the pump laser, with a center wavelength ($\lambda_c$) of 405.1 nm and an FWHM ($\Delta \lambda$) of 0.53 nm.  (b) is the spectrum of the signal, with $\lambda_c$ = 810.3 nm and $\Delta \lambda$ = 5.57 nm. (c) is the spectrum of the idler, with $\lambda_c$ = 809.7 nm and $\Delta \lambda$ = 7.32 nm. In (a, b, c), each blue line corresponds to experimental data, and each red line is a Gaussian fit. (d) is the simulated JSI of the biphotons.
 } \label{fig:2}
\end{figure}
The experimental setup is shown in Fig.\,\ref{fig:1}.
A broadband multi-mode LD was utilized as the pump laser.
A typical spectrum of the LD is shown in Fig.\,\ref{fig:2}(a), with a central wavelength of 405.1 nm and a full-width-at-half-maximum (FWHM) of 0.53 nm, which was measured by a spectrometer of Princeton Instruments SP2300.
The pump laser was coupled into a single-mode fiber (SMF) to filter the spatial mode, and the SMF with an FC/APC-type connector also functioned as an isolator to block the back-reflected pump laser from the Sagnac loop.
After passing through a QWP (quarter-wave plate), an HWP (half-wave plate), an SPF (short-wavelength pass filter),  and a lens (L$_1$), the pump photons were sent into a triangle-shape Sagnac loop, which had a compact size of about 9 cm + 12.7 cm + 9 cm.
In the Sagnac loop, the DPBS (dual-wavelength polarization beam splitter) and the DHWP (dual-wavelength HWP) worked for both 405 and 810 nm wavelengths. The PPKTP crystal was type-II phase-matched (y$\to$y+z) with a poling period of 9.825 $\mu$m.
The downconverted biphotons, i.e., the signal (y-polarized) and the idler (z-polarized), had a degenerate wavelength of around 810 nm at the temperature of 92.5 $^{\circ}$C, and their FWHMs were measured to be 5.57 nm and 7.32 nm, respectively, as shown in Fig.\,\ref{fig:2}(b-c).
We also theoretically simulated their joint spectral intensity (JSI), as shown in Fig.\,\ref{fig:2}(d) (Also see Fig. A1 in the Appendix). The simulated JSI explained the experimental phenomenon that the signal has a narrower bandwidth than the idler. For simplicity, in this simulation, we assumed the PPKTP crystal was 10 mm long, and the spectrum of the pump laser had a Gaussian distribution with a center wavelength of 405 nm and a bandwidth of 0.45 nm.
Under this condition, the simulated FWHMs of the signal and idler are 5.25 nm and 7.78 nm, which are in good agreement with the experimental results.
The biphotons generated from the clock-wise pump and the counter-clock-wise pump were separated by a DPBS, collimated by two lenses (L$_2$ and L$_3$ ), filtered by two LPF (long pass filters), adjusted by two polarizers, and finally coupled into two SMFs, which was connected to single-photon detectors (SPCM-AQRH-10-FC, Excelitas) and a coincidence counter (Picoharp 300, PicoQuant).
For quantum correlation measurement, each polarizer was composed of an HWP and a PBS.

\begin{figure}[!tb]
\centering\includegraphics[width= 0.48 \textwidth]{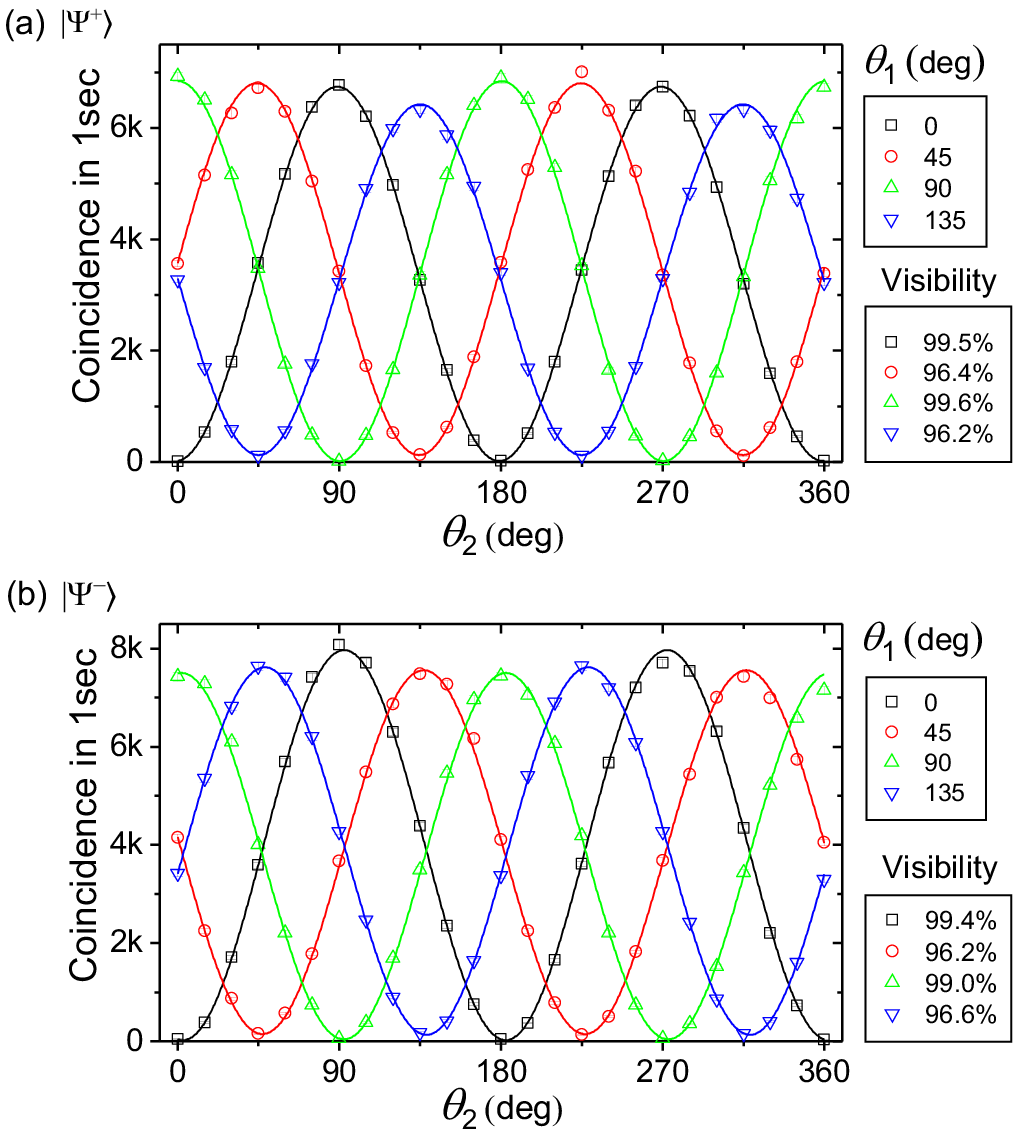}
\caption{Polarization correlation measurement for $|\Psi^+\rangle$ (a) and  $|\Psi^-\rangle$ (b).
$\theta_{1(2)}$ is the angle of polarizer$_{1(2)}$. The error bars were added by assuming Poissonian statistics of these coincidences counts.
 } \label{fig:3}
\end{figure}

\begin{figure}[!tbh]
\centering\includegraphics[width= 0.48 \textwidth]{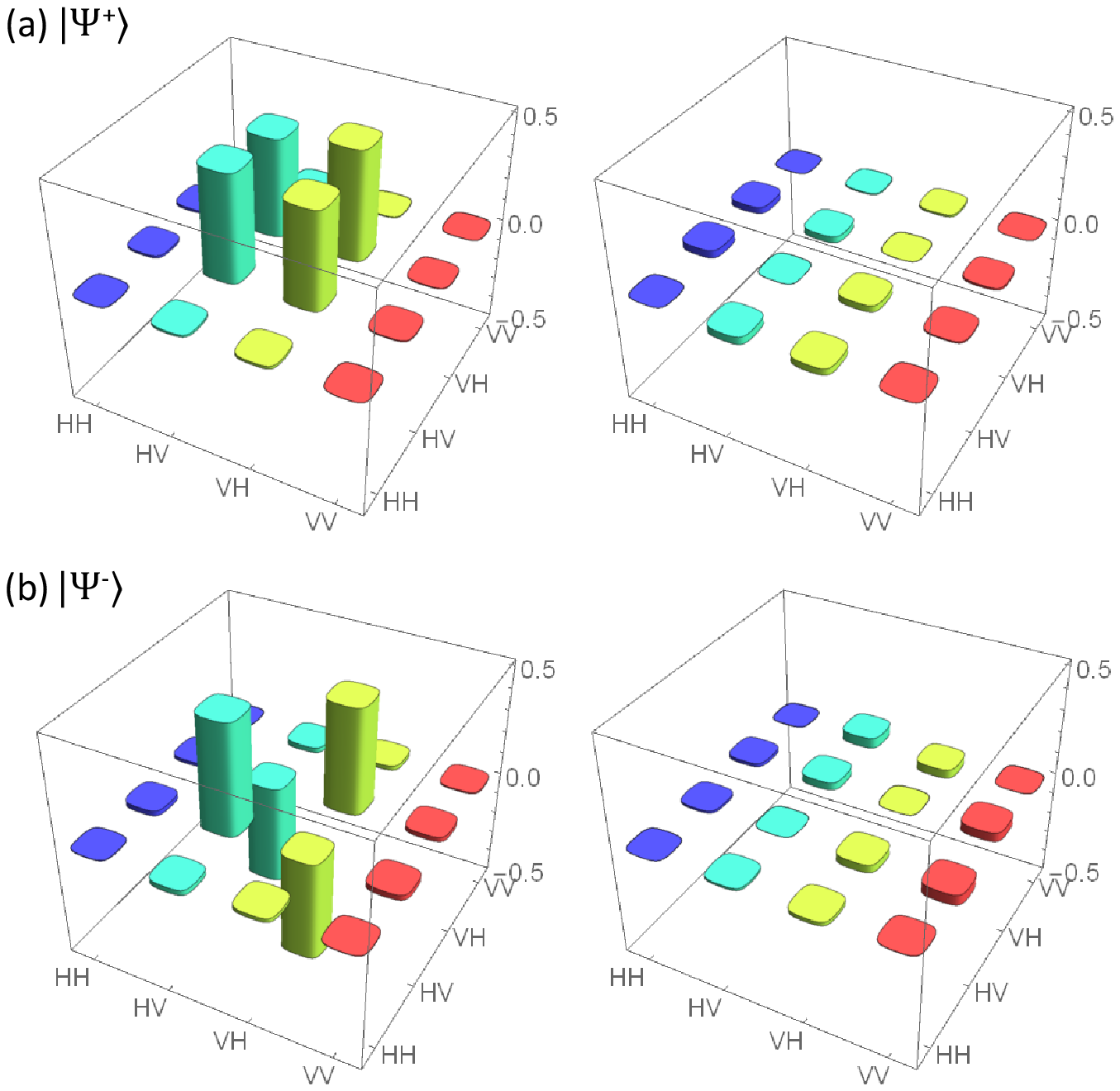}
\caption{Real (left) and imaginary (right) part of the reconstructed density matrix obtained by the maximum likelihood estimation method.(a) for \(|\Psi^+\rangle\) state;(b) for \(|\Psi^-\rangle\) state.} \label{fig:4}
\end{figure}

%
As theoretically analyzed in Ref. \cite{Kim2006}, the output state from the PPKTP-Sagac scheme is
\begin{equation}\label{eq1}
|\Psi\rangle \propto (|H\rangle|V\rangle+e^{i\phi}\beta|V\rangle|H\rangle),
\end{equation}
where $\phi$ and $\beta$ are the relative phase and pump ratio between the clock-wise path and counter-clock-wise  path.
By rotating HWP, QWP and by finely adjusting the position of PPKTP, we can prepare the Bell states $\left| {\Psi ^ +  } \right\rangle  = \frac{1}{{\sqrt 2 }}(\left| {HV} \right\rangle {\rm{ + }}\left| {VH} \right\rangle )$ and $ \left| {\Psi ^ -  } \right\rangle  = \frac{1}{{\sqrt 2 }}(\left| {HV} \right\rangle  - \left| {VH} \right\rangle )$.
Figure \ref{fig:3} shows the polarization correlation measurement result for  $|\Psi^\pm\rangle$ states.
For $|\Psi^+\rangle$ state,  the maximal coincidence was around 7 kcps with a pump power a 7 mW at each projection base. The corresponding overall brightness was 2 kcps/mW.
Without any background subtraction, the visibilities of 0$^{\circ}$, 45$^{\circ}$, 90$^{\circ}$, and 135$^{\circ}$ were 99.5\%, 96.4\%, 99.6\%, 96.2\%, respectively.
For $|\Psi^-\rangle$ state, the corresponding visibilities were 99.4\%, 96.2\%, 99.0\% and 96.6\%, respectively.
All the fringe visibilities in Fig.\,\ref{fig:3} were higher than 71\%, the bound
required to violate the Bell's inequality.
We also measured the Bell parameter S, which directly indicated the violation of Bell's inequality.
The S values measured were 2.78$\pm$0.01 and 2.74$\pm$0.01 for $|\Psi^+\rangle$ and $|\Psi^-\rangle$, respectively, which were higher than the classical bound of 2.

We also performed the quantum state tomography of the polarization-entangled state.
Polarizers 1 and 2 in  Fig.\,\ref{fig:1} were replaced by combinations of HWP, QWP, and PBS.
The density matrix was reconstructed with a maximum likelihood estimation method, as shown in  Fig.\,\ref{fig:4}.
For states \(|\Psi^+\rangle\) and \(|\Psi^-\rangle\), the fidelities of the reconstructed density matrix were 0.948$\pm$0.004 and 0.963$\pm$0.002, respectively.
These values indicated that the states were highly entangled.

\section{Discussion}
We compare this work with the previous single-mode LD pumped Sagnac-PPKTP scheme used for satellite application \cite{Yin2017}, the multi-mode laser pumped ``universal Bell-state synthesizer'' scheme \cite{Jeong2016}, and the multi-mode laser pumped ``linear beam displacement interferometer'' scheme \cite{Lohrmann2020APL} in Tab.\,\ref{Tab:1}.
It is noteworthy that the type-0 phase-matched crystal has the highest brightness because the nonlinear coefficient of $d_{33}$ is much higher than the value of $d_{24}$ in type-II phase-matched condition. However, the spectral widths of the biphotons in type-0 phase-matched case are much broader.
The Sagnac-PPKTP scheme can achieve comparable performance as the ``universal Bell-state synthesizer'' \cite{Kim2003} scheme in stability and brightness, but the configuration is more compact.
The single-mode LD laser pumped Sagnac-PPKTP scheme can also achieve high performance, but the pump laser has a higher cost.

\begin{table*}[!tb]
\centering
\resizebox{\linewidth}{!}{
\begin{tabular}{c|c|c|c|c}
\hline \hline
Parameter    & Yin2017 \cite{Yin2017}   & Jeong2016 \cite{Jeong2016}   & Lohrmann2020  \cite{Lohrmann2020APL}   & This work\\
\hline
Configuration  & Sagnac loop
             & \tabincell{c}{universal Bell-state\\synthesizer}
             &\tabincell{c}{linear beam displacement\\ interferometer}
             & Sagnac loop \\
\hline
 \tabincell{c}{Pump laser \\ $\lambda_c$ and $\Delta \lambda$}   & \tabincell{c}{Single-mode LD\\405 nm, 160 MHz}
             & \tabincell{c}{Multi-mode LD\\ 406.2 nm, 0.5 nm}
             & \tabincell{c}{Multi-mode LD\\ 405.5 nm, $\approx$0.5 nm}
             & \tabincell{c}{Multi-mode LD\\ 405.1 nm, 0.53 nm} \\
\hline
PPKTP        & \tabincell{c}{L=15 mm, --  \\ Type-II, collinear}
             & \tabincell{c}{L=10 mm, $\Lambda$=10 $\mu$m \\ Type-II, non-collinear}
             & \tabincell{c}{L=10 mm, $\Lambda$=3.425 $\mu$m \\ Type-0, collinear}
             & \tabincell{c}{L=10 mm, $\Lambda$=9.825 $\mu$m \\ Type-II, collinear}\\
\hline
\tabincell{c}{Biphotons \\ $\lambda_c$ and $\Delta \lambda$}   & \tabincell{c}{811 nm,\\ -- }
             & \tabincell{c}{812.4 nm, \\5.8/8.7 nm }
             & \tabincell{c}{780/842 nm,\\ over 100 nm }
             & \tabincell{c}{810 nm, \\  5.6/7.3 nm} \\
\hline
\tabincell{c}{Brightness \\ (Kcps/mW) }   & \tabincell{c}{197 \\ (5.9 Mcps/30 mW)}
             & \tabincell{c}{SMFs: 7  \\MMFs: 90.9 }
             & 560
             & 2 \\
\hline
Visibility   & over 91\%
             & \tabincell{c}{SMFs: 97.8\% \\MMFs: 93.6\%}
             & over 96.4\%
             & over 96.2\%\\
\hline
Fidelity     & \tabincell{c}{0.907 $\pm$ 0.007 \\ (at 5.9 Mcps)}
             & \tabincell{c}{SMFs: 0.992 \\MMFs: 0.968}
             & --
             & \tabincell{c}{$ \left|{\Psi ^ -}\right\rangle$: 0.963 \\$ \left|{\Psi ^ +}\right\rangle$: 0.948}\\
\hline \hline
\end{tabular}
}
\caption{Comparison of this work with the previous results. L is the length of the PPKTP crystal,  and $\Lambda$ is the poling period.  $\lambda_c$ is the center wavelength,  and $\Delta \lambda$ is the FWHM.}
\label{Tab:1}
\end{table*}

The models and characteristics of the components used in this work are listed in Tab. \ref{Tab:A1} in the Appendix.
It can be noticed that the entangled source used in this experiment can still be optimized by improving the collection efficiency of the whole system, especially the coupling efficiency to the SMF, the transmission efficiency of the LPFs.
For example, we can optimize the beam waist of the pump laser by using a proper focusing lens and choose LPF with higher transmission efficiency.
Nevertheless, we have shown that it is possible to prepare a highly entangled photon source using multi-mode LD pumped PPKTP-Sagnac configuration.

For future applications this highly entangled photon source is applicable not only for foundational tests of quantum physics but also for quantum networks, e.g., entanglement-based quantum key distribution networks \cite{Liu2020APLZhangwei}.  As shown in Fig.\,\ref{fig:2}(d), the signal and idler photon are also entangled in frequency, so this source has the potential to be a hyper-entangled photon source \cite{Ramelow2009PRL, YuanyuanChen2020}.  Further, the feature of low-cost makes it applicable for educational use. But this source is not useful for Hong-Ou-Mandel (HOM)-interference-based applications, e.g., teleportation or entanglement swapping, because the HOM interference visibility is very low due to the low spectral exchanging symmetry.

\section{Conclusion}
In summary, we have demonstrated the combination of a broadband LD,  a PPKTP,  and a Sagnac loop to generate polarization-entangled photons.
In polarization correlation measurement, the visibilities are all over 96\%, and the S value of the Bell's inequality reached 2.78$\pm$0.01.
In a quantum state tomography measurement, the fidelity achieved 0.963$\pm$0.002.
This entangled source has the merits of being cost-effective, compact, high-brightness, and high stability and may provide a good option for practical applications in quantum information processing.

\section*{Acknowledgement}
This work is supported by the National Natural Science Foundations of China (Grant Nos.12074299, 91836102, 11704290) and by the Guangdong Provincial Key Laboratory (Grant No. GKLQSE202102).




%

\renewcommand\thefigure{A\arabic{figure}}
\setcounter{figure}{0}

\renewcommand\thetable{A\arabic{table}}
\setcounter{table}{0}

\setcounter{equation}{0}
\renewcommand\theequation{A\arabic{equation}}


\section*{Appendix}
\begin{figure*}[!htb]
\centering\includegraphics[width= 0.98 \textwidth]{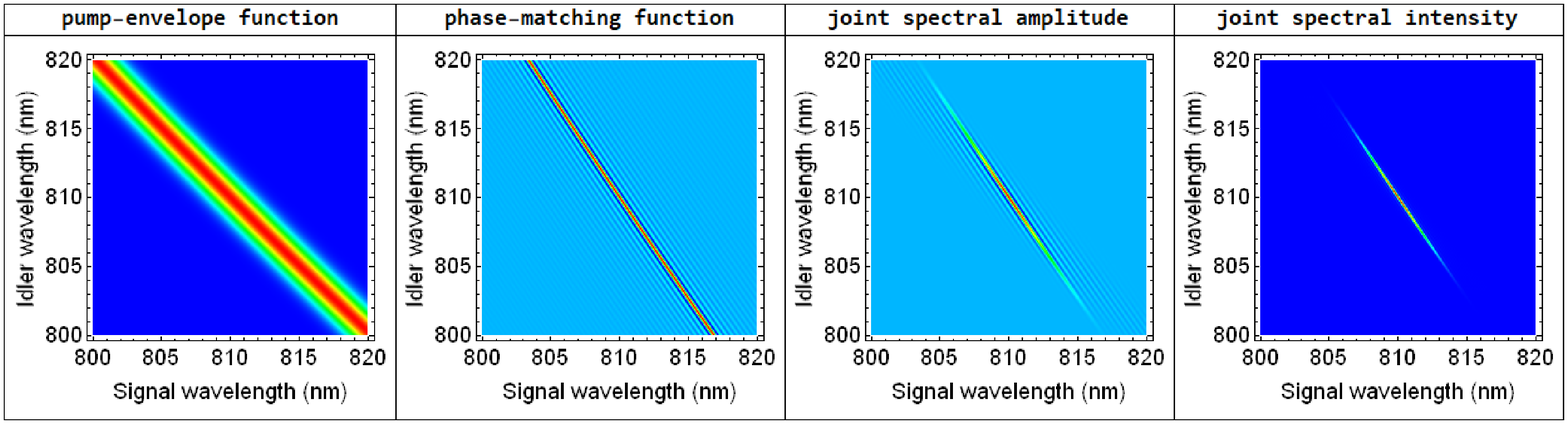}
\caption{The joint spectral amplitude (JSA) is the product of pump-envelope function (PEF) and phase-matching function (PMF) \cite{Jin2013OE}, and the absolute square of JSA is JSI, which is shown in Fig. 2(d).} \label{fig:A1}
\end{figure*}

Figure\,\ref{fig:A1} shows how the JSI in Fig.\,2(d) was simulated.

The models and performance of the experimental components in this work are shown in Tab.\,\ref{Tab:A1}.

\begin{figure*}[!htp]
\centering
\includegraphics[width= 0.88 \textwidth]{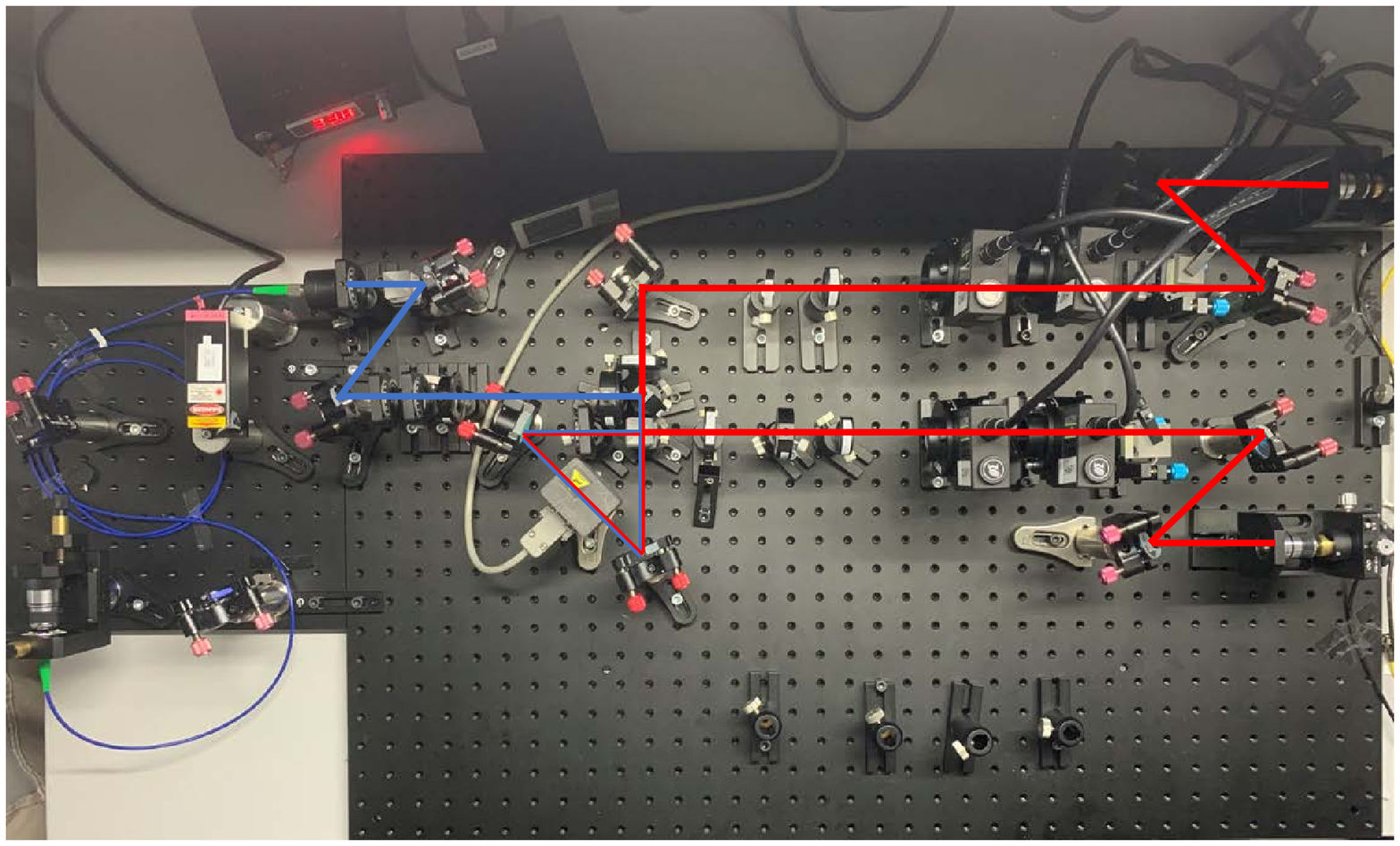}
\caption{A photograph of the setup.} \label{fig:A2}
\end{figure*}

A photograph of the setup is shown in Fig.\,\ref{fig:A2}.

\begin{table*}[!htb]
\centering
\resizebox{\linewidth}{!}{
\begin{tabular}{ l l l}
\hline \hline
Name & Type & Characteristics \\
\hline
Laser& LR-BSP-405nm/100mw (LR laser Co.) & $\lambda_c$=405.1 nm, FWHM=0.53 nm\\
Fiber coupler & GCX-L005-FC (Daheng Co.) & $\eta_c$=52.6\%\\
SMF (405 nm)& PM-S405-XP (Thorlabs) &  with FC/APC connector\\
Mirror (405 nm)& LLM0025-45-397-405 (Union Optics Co.) & $\eta_r$=99.2\%\\
QWP (405 nm)& WPZ4420-405 (Union Optics Co.)& $\eta_t$=100\%\\
HWP (405 nm) & WPZ2420-405 (Union Optics Co.)&$\eta_t$=100\%\\
PBS (405 nm)& PBS0120-397-405 (Union Optics Co.) & extinction ratio=1000:1\\
Lens 1 (405 nm) & PCX1809-300-500 (Union Optics Co.) & f=200 mm, $\eta_t$=100\%; fused silica\\
Lens 2/3 (810nm)& PCX0810-780-810 (Union Optics Co.)& f=200 mm, $\eta_t$ =100\%; K9 glass\\
Mirror (810 nm) & GCC-102202 (Daheng Co.) & protected silver with $\eta_r$=98.2\%\\
QWP (810 nm)& GCL-060704 (Daheng Co.)& zero order QWP\\
HWP (810 nm) & GCL-060714 (Daheng Co.) & zero order HWP\\
SPF& GCC-211002 (Daheng Co.) &  $\eta_t$=96.7\% for 400-630 nm \\
LPF1(DM) & DIM-K9-25.4-3 (Union Optics Co.)& $\eta_t$=96\%(810 nm),$\eta_r$=94\%(405 nm)\\
LPF2& RG-715 (Edmund Co.) & $\eta_t$=89.7\%(810 nm)\\
DHWP & DHWP (Union Optics Co.) & HWP for 405 nm and 810 nm\\
DPBS& DPBS (Union Optics Co.) & 405nm: $\eta_t$=93.1\%, $\eta_r$=97.4\%\\
 & &810nm: $\eta_t$=96.2\%, $\eta_r$=96.7\%\\
SMF (810 nm) & GCX-XSM-4/125-FC/PC (Daheng Co.) & SMF at 810 nm \\
PPKTP& PPKTP (Raicol Co.)& size=1$\times$2$\times$10 mm, $\Lambda$=9.825 $\mu$m\\
Oven & TC038-PC (HC Photonics Co.) &  resolution=0.1$^{\circ}$C, range=0-200$^{\circ}$\\
Rotator& OSMS-60YAW (Sigma Koki Co. ) & controlled by LabVIEW software\\
Si-APD   & SPCM-AQRH-10-FC (Excelitas Co. )&  $\eta_d \approx$ 60\% at 810 nm \\
\hline \hline
\end{tabular}
}
\caption{\label{Tab:A1} The models and characteristics of the main components for the LD-pumped PPKTP-Sagnac entangled photon source.
$\eta_c$ is the coupling efficiency, $\eta_d$  is the detection efficiency, $\eta_t$ is the transmission ratio,   and $\eta_r$ is the reflection ratio.}
\end{table*}

\end{document}